\documentclass{article}%
\usepackage{graphicx}
\usepackage{amsmath}%
\usepackage{amsfonts}%
\usepackage{amssymb}
\newtheorem{theorem}{Theorem}
\newtheorem{acknowledgement}[theorem]{Acknowledgement}

\begin{document}

\title{Modeling Nonlinear Dynamical Systems with Delay-differential Equations. }
\author{Alexander N. Jourjine\\Electrical Engineering Dept\\Princeton University\\Princeton 08540, USA\\jourjine@princeton.edu}
\date{December 24, 2000}
\maketitle

\begin{abstract}
We describe a method to model nonlinear dynamical systems using periodic
solutions of delay-differential equations. We show that any finite-time
trajectory of a nonlinear dynamical system can be loaded approximately into
the initial condition of a linear delay-differential system. It is further
shown that the initial condition can be extended to a periodic solution of the
delay-differential system if an appropriate choice of its parameters is made.
As a result, any finite set of trajectories of a nonlinear dynamical system
can be modeled with arbitrarily small error via a set of periodic solutions of
a linear delay-differential equation. These results can be extended to some
non-linear delay differential systems. One application of the method is for
modeling memory and perception.

\end{abstract}

\section{Introduction}

How the information about the outside world is stored in the brain is still in
many ways an open question. Clearly, representation of the totality of the
information involves some sort of modeling of the environment by the brain
under the specific conditions of its operation, e.g., the limited number of
the neurons and synapses, the inherent delays in signal propagation along the
neural fibers, \textit{etc}. Assuming that the environment is described by
some nonlinear deterministic dynamical system and that the brain is described
by another dynamical system, this question can be rephrased as the question of
how one dynamical system can model another. Since the environment is typically
much more complicated than the brain, the modelling task should be impossible
simply on the count of the difference in the number of the degrees of freedom
for the two dynamical systems. A possible solution to this puzzle is to assume
that the environment is actually a collection of weakly interacting
subsystems, each of which has a lower complexity than the brain. A good
strategy to learn complicated environment would then be to model the strongly
coupled degrees of freedom group by group. Decomposition of the environment
into a collection of components that interact weakly presumably should involve
some version of nonlinear factor analysis extended to deterministic systems.

Even if the complexity of the environment is less than that of the system that
models it a fundamental question arises about in what sense and how one
dynamical system can model another. In addition one can ask whether there
exist a class of dynamical systems that in some sense are universal in their
modeling properties in the sense that a large class of nonlinear dynamical
systems that represent possible environments can be modeled using essentially
the same system. Of course, in order for this to make sense in explaining the
properties of the brain, the modeling has to be physically realizable.

Leaving investigation of the deterministic nonlinear factor analysis to
another publication \cite{Jourjine_Factor}, in this paper we present arguments
that when the dimension of the environment is less or equal to that of the
modeling system such universal modeling systems exist and that it is by the
essential use of the delays that the universality can be realized. We prove
that under certain conditions any nonlinear dynamical system can be modeled by
a linear delay-differential dynamical system with constant coefficients, in
the sense that any finite set of trajectories of the environment can be loaded
into periodic solutions of the delay-differential equation. The loading turns
out to be equivalent to solving an essentially linear problem and is achieved
by adjusting the coefficients of the delay-differential equation in a way that
is plausible from the biological point of view. With the biological
applications in mind we also extend some of the results to a class of
nonlinear systems with delays.

Modeling the environment is usually referred to as construction memory models
of the environment. Since the early eighties much work has been done on memory
models under the assumption that time-independent memories can be represented
as fixed points of systems of nonlinear differential equations (ODE) that
possess a Liapunov function and with the evolution given by%
\begin{equation}
\dot{x}\left(  t\right)  =F\left(  x\right)  ,\ x(t)\in R^{N} \label{nl_ODE}%
\end{equation}
for some function\ $F\left(  x\right)  $. Once a Liapunov function is given,
the fixed points of the evolution can be identified with the $N$-dimensional
stored patterns. Domains of attraction of the fixed points then can be viewed
as all the patterns that will be ''associatively'' recalled dynamically as the
stored pattern represented by the fixed points. Especially useful within this
approach proved to be the Hopfield model \cite{Hopfield} described as a
dynamical system with evolution
\begin{equation}
\dot{x}\left(  t\right)  =-E_{\mu}\cdot x\left(  t\right)  +A\cdot
\sigma(x\left(  t\right)  )+y, \label{Hopfield}%
\end{equation}
where $E_{\mu}=diag\{\mu_{i}\},\ \mu_{i}>0$ is an $N\times N$ diagonal
matrix$,$ $A$ - is an $N\times N$ matrix, called the matrix of weights,
$\sigma:x_{i}\rightarrow\sigma_{i}(x_{i})$ is a component-wise nonlinear
transformation and the constant vector $y\in R^{N}$ describes time-independent
external inputs to the system. If $\sigma^{\prime}(z)\geq0$ and matrix $A$ is
symmetric, or if the symmetric part of $A$ is non-positive definite
\cite{Jourjine}, then the system $\left(  \ref{Hopfield}\right)  $ has a
Liapunov function $V\left(  x\right)  $ defined on the configuration space
such that on the trajectories $\dot{V}(x(t))<0,$ leading to the convergence of
all trajectories to a set of fixed points. An extensive theory exists about
the properties of memory storage, especially for random patterns
\cite{McEliece}.

A certain amount work has been done for the systems with delays. Existence of
a Liapunov function can be proven for the Hopfield model with delays if the
delays are not too large \cite{Ma}. However, the fixed-points paradigm is
difficult to apply to the time dependent patterns. Time-variable patterns,
which in our approach are represented by the trajectories of nonlinear
dynamical systems, are ubiquitous in the environment. Therefore, understanding
the underlying principles of their storage is important when constructing
plausible models for perception and memory. Some interesting effects of the
presence of delays on the structure of the attractors leading to
multistability in networks of two Hodgkin-Huxley type spiking neurons were
discussed heuristically and numerically in \cite{Foss}. From our point of view
the model that is considered there is that of a system of two linear delay
differential equations with time variable coefficients. In this paper we do
not restrict ourselves to the models of spiking neurons but consider
arbitrarily large systems of linear and nonlinear delay-differential equations
with slowly varying coefficients. Systems with time variable coefficients will
be considered in detail elsewhere. Nevertheless, our results are in general
agreement with the numerical simulations in \cite{Foss} and provide a new
perspective on their interpretation.

Here we present an approach to the time-dependent pattern storage that
describes sets of interacting ''neurons'' as systems of delay-differential
equations (DDE) with multiple delays. Since delays in nerve impulse
propagations are common, the approach is more plausible biologically than, for
example, the Hopfield model where instantaneous communication among the
neurons is assumed. Mathematically the delays appear after integrating out
some internal degrees of freedom in systems of Hodgkin-Huxley equations
\cite{Hodgkin}. Although more complicated to analyze, the DDE have some
properties that ordinary differential equations do not have. For example, for
a solution of a DDE to be unique, one has to specify an initial condition
which is a function defined on a finite time range \cite{Bellman}. This is a
key feature of DDE that we exploit.

Our starting point is\ to include explicitly the multiple delays $\tau
_{1},...,\tau_{L}$ in the dynamics of memory model so that instead of $\left(
\ref{nl_ODE}\right)  $ we obtain a system of nonlinear DDE%
\begin{equation}
\dot{x}\left(  t\right)  =F\left(  x\left(  t\right)  ,x\left(  t-\tau
_{1}\right)  ,...,x\left(  t-\tau_{L}\right)  \right)  . \label{nl_DDE}%
\end{equation}
One example of such a system is a generalization of the Hopfield model to
include delays. This was considered in \cite{Ma}%
\begin{equation}
\dot{x}\left(  t\right)  =\sum_{k=0}^{L}A_{k}\cdot\sigma\left(  x\left(
t-\tau_{k}\right)  \right)  +y(t), \label{extend_hopf}%
\end{equation}
where as before the $A_{k}$ and are $N\times N$ weight matrices with
$A_{0}=-E_{\mu},$ and the delays are defined so that $\tau_{0}=0,\tau_{k}$
$\leq$ $\tau_{k+1}$. In contrast to $\left(  \ref{Hopfield}\right)  $ we
assume the external inputs $y(t)$ as time-variable.\ In models describing
networks of oscillating neurons one can also consider $\tau_{k}$ as an
$N\times N$ matrix with each element $\left(  \tau_{k}\right)  _{ij}$
describing the delay in signal propagation from neuron $\left(  j\right)  $ to
neuron $\left(  i\right)  $ via $"k"$th pathway$.$ However, after appropriate
redefinition of matrices $A_{k}$ and nonlinearities $\sigma\left(  z\right)  $
one can recover $\left(  \ref{extend_hopf}\right)  $ at the expense of
increasing $L$ to the number of different values of all matrix elements
$\tau_{k}$ of all $\left(  L+1\right)  $ matrices. We assume that $y(t)$ is a
trajectory of some nonlinear dynamical system with local evolution\ given by
$\dot{y}(t)=G(y).$ The detailed nature of $G(y)$ is not important for the
discussion. When $\sigma_{i}(z)=z$ we obtain a system of linear DDE. Writing
it in components we obtain%
\begin{equation}
\dot{x}_{i}=\sum_{k=0}^{L}\left(  A_{k}\right)  _{ij}\cdot x_{j}\left(
t-\tau_{k}\right)  +y_{i}(t),\;i=1,...,N. \label{linear_DDE}%
\end{equation}

The main point of this paper is to prove the existence of the periodic
solutions of systems of homogeneous linear DDE that are periodic extensions of
their initial conditions. That is, we show that, given a set of fixed delays
$\tau_{k}$, the weight matrices $A_{k}$ can be chosen in such a way that the
initial condition when extended periodically to $t>\tau_{L}$ is the solution
of $\left(  \ref{linear_DDE}\right)  $. With appropriate choice of parameters
the choice of $A_{k}$ is unique. We also show that if the inhomogeneous part
of a DDE is given by a trajectory of a nonlinear dynamical system, the
trajectory can be loaded approximately, but with arbitrarily small error into
the initial condition for the DDE and hence stored as its periodic solution.
With some modifications these results can be carried over to a class of
nonlinear DDE. Using the method we describe, with a sufficiently large $N,$
any finite set of trajectories of a nonlinear dynamical system, for example
any finite set of periodic orbits, can be encoded as periodic solutions of a
linear DDE. Since many dynamical systems, are uniquely characterized by the
sets of their periodic orbits, we can speak of linear modelling of nonlinear systems.

The paper is structured as follows. In the next section we give a brief
overview of the differences between the linear ODE and the linear DDE from
mathematical point of view. Section 3 contains discussion of loading the
trajectories into the initial conditions. In Section 4 we construct solutions
for the periodic extension problem. In Section 5 we discuss the adaptation
dynamics for the weight matrices $A_{k}$. Section 6 is a summary.

\section{Delay-differential Equations}

To specify a solution of a differential equation uniquely one needs to pose
the initial value problem. There is a big difference in how the initial value
problem is posed for ODE and for DDE. For the ODE the initial value problem is
given by

\begin{description}
\item $\qquad\qquad x\left(  \phi,y\right)  \left(  t\right)  $ is a solution
of $\left(  \ref{nl_ODE}\right)  $ for $t>0;$

\item $\qquad\qquad x\left(  \phi,y\right)  \left(  0\right)  =\phi\in R^{N}.$
\end{description}

For the DDE of retarded type $\left(  \ref{nl_DDE}\right)  $ with the maximum
delay $\tau_{L}$ the initial value problem is formulated as

\begin{description}
\item $\qquad\qquad x\left(  \phi,y\right)  \left(  t\right)  $ is a solution
of $\left(  \ref{nl_DDE}\right)  $ for $t>\tau_{L};$

\item $\qquad\qquad x\left(  \phi,y\right)  \left(  0\right)  =\phi\left(
t\right)  $ $\in R^{N}$ for $t\in\left[  0,\tau_{L}\right]  .$
\end{description}

Hence to specify a unique solution of a DDE one needs an initial value from a
space of functions defined on the interval $\left[  0,\tau_{L}\right]  $.
Exploiting the difference is one key feature of our approach.

Let us consider in more detail the difference in the linear case. Similar to
homogeneous linear ODE we can search for the exponential solutions of
homogeneous DDE. As for ODE these correspond to zeroes of the characteristic
equation that is obtained using the Laplace transform. For linear ODE the
characteristic equation is polynomial and has exactly $N$ complex roots. The
situation is different for linear DDE. The characteristic equation becomes
transcendental and is given by%
\begin{align}
\det\left(  H\left(  \lambda\right)  \right)   &  =0,\label{char_eq}\\
H\left(  \lambda\right)   &  =\lambda\cdot I-\sum_{k=0}^{L}A_{k}\exp\left(
-\lambda\tau_{k}\right)  .\nonumber
\end{align}
There are generically infinite number of complex roots $\lambda_{p}$ of
$\left(  \ref{char_eq}\right)  .$ Let $N(a,b),$ $a,b\in R,$ $a\leq b<\infty,$
be the number of the complex roots of $\left(  \ref{char_eq}\right)  $ in the
vertical strip of the complex plain with real part of each point contained in
$\left[  a,b\right]  .$ Then, some of useful properties of the roots are\qquad

\begin{description}
\item \qquad\qquad(i) Re$\left(  \lambda_{p}\right)  \leq c<\infty$ if all
$\tau_{k}\geq0,$ for all $k,i,j;$

\item \qquad\qquad(ii) Re$\left(  \lambda_{p}\right)  \rightarrow-\infty$
\ \qquad$p\rightarrow\infty;$

\item \qquad\qquad(iii) $N(a,b)<\infty$.
\end{description}

Solutions of linear \textit{inhomogeneous} DDE can be conveniently written
down in an integral form using the notion of the fundamental solution. The
fundamental solution is defined as the unique matrix solution of $\left(
\ref{linear_DDE}\right)  $ with $y\left(  t\right)  =0$ and with the following
initial condition

\begin{description}
\item $\qquad\qquad X\left(  t\right)  =0$ \ $t\in\left[  0,\tau_{L}\right]  ;$

\item $\qquad\qquad X\left(  t\right)  =I$ \ $t=\tau_{L},$ where $I$ is the
unit matrix.
\end{description}

Using inverse Laplace transform, the fundamental solution can be expressed in
terms of the characteristic (matrix) function $H\left(  \lambda\right)  $%
\begin{equation}
X\left(  t\right)  =\frac{1}{2\pi i}\int_{\left(  c\right)  }d\lambda
\;e^{\lambda t}H^{-1}\left(  \lambda\right)  , \label{fund_sol}%
\end{equation}
where $\left(  c\right)  $ is the contour $\left(  c-i\infty,c+i\infty\right)
$ in the complex plane such that $\operatorname{Re}\left(  \lambda_{p}\right)
<c$ and where the Laplace transform and its inverse are defined by%
\begin{align}
\hat{x}\left(  \lambda\right)   &  =\int_{0}^{\infty}d\lambda\;\exp\left(
-\lambda\ t\right)  \ x\left(  t\right)  ,\label{Laplace_tr}\\
x\left(  t\right)   &  =\frac{1}{2\pi i}\int_{\left(  c\right)  }%
d\lambda\;\exp\left(  -\lambda\ t\right)  \ \hat{x}\left(  \lambda\right)  .
\label{Laplace_inv_tr}%
\end{align}
With these definitions the unique solution of the initial value problem for
$\left(  \ref{linear_DDE}\right)  $ can be written as \cite{Bellman}%
\begin{align}
x\left(  \phi,y\right)  \left(  t\right)   &  =X\left(  t-\tau_{L}\right)
\phi\left(  \tau_{L}\right)  +\int_{\tau_{L}}^{t}d\tau\;X\left(
t-\tau\right)  \;y\left(  \tau\right) \label{sol_DDE_fund}\\
&  +\sum_{k=1}^{L}\int_{\tau_{L}-\tau_{k}}^{\tau_{L}}d\tau\;X\left(
t-\tau-\tau_{k}\right)  \;A_{k}\;\phi\left(  \tau\right)  .\nonumber
\end{align}
If we put $L=0$ then the third term in $\left(  \ref{sol_DDE_fund}\right)  $
vanishes and we recover the solution of a linear inhomogeneous ODE in terms of
its fundamental solution. Note that in $\left(  \ref{sol_DDE_fund}\right)  $
the inhomogeneous part $y\left(  t\right)  $ and the initial condition
$\phi\left(  t\right)  $ play a similar role. We shall exploit the similarity
below when we generate initial conditions from the inhomogeneous part of the equation.

\section{Loading Trajectories into the Initial Conditions}

In this section we describe how to generate the initial condition for a DDE
from its inhomogeneous part. The main idea is to consider the inhomogeneous
part to be nonzero only on $\left[  0,\tau_{L}\right]  $ and generate the
initial condition recursively by adding more and more terms to the equation successfully.

Let us first consider the linear DDE. We divide the interval $\left[
0,\tau_{L}\right]  $ into adjoining intervals $\left[  \tau_{k,}\tau
_{k+1}\right]  ,\ k=0,...,L-1$ and for time $t<$ $\tau_{m}$ \ truncate the
full DDE to%
\begin{equation}
\dot{x}\left(  t\right)  =\sum_{k=0}^{m}A_{k}\cdot x\left(  t-\tau_{k}\right)
. \label{linear_DDE_trunc}%
\end{equation}
This can be done under the assumption that the signals with larger delays did
not arrive yet at the ''neuron'' to influence its state.

When $m=0$ $\left(  \ref{linear_DDE_trunc}\right)  $ reduces to a linear ODE
which can be integrated for $t\in(0,\tau_{1}]$ using the initial condition
$\phi_{0}$ at $t=0$ and the values of the inhomogeneous part $y(t)$. The
result of the integration can be considered as the initial condition the
truncated DDE defined on $t\in(\tau_{1},\tau_{2}]$, which enables us to
integrate the DDE \ with one delay $\tau_{1}$on the segment $\left[  \tau
_{1,}\tau_{2}\right]  $ using $y(t)$ defined on $\left[  \tau_{1,}\tau
_{2}\right]  $ only. Proceeding in this fashion step by step we can construct
the initial condition on the entire segment $\left[  0,\tau_{L}\right]  $.
\ Taking the inhomogeneous part $y(t)$ to be zero outside of $\left[
0,\tau_{L}\right]  $ we can interpret the result of this procedure as a linear
mapping of trajectories into the initial conditions, $I_{L}:$ $y(t)$
$\rightarrow\Phi\left(  t\right)  ,t\in$ $\left[  0,\tau_{L}\right]  $.

To describe the iteration procedure in more detail we give the initial
iteration step followed by the $"m"$th iteration. For the initial step we need
to generate the initial condition for%
\begin{align}
\dot{x}\left(  t\right)   &  =A_{0}\cdot x\left(  t\right)  +A_{1}\cdot
x\left(  t-\tau_{1}\right)  +y(t),\label{step_1_DDE}\\
y(t)  &  =0,t\notin\left[  0,\tau_{1}\right]  .\nonumber
\end{align}
These are given by the solution of the initial value problem%
\begin{align}
\dot{x}\left(  t\right)   &  =A_{0}\cdot x\left(  t\right)
+y(t),\label{first_iter_init_val_prob}\\
x(0)  &  =\phi_{0}.\nonumber
\end{align}
Using the fundamental solution for this equation the solution\ for $t>0$ and
the initial condition for $\left(  \ref{step_1_DDE}\right)  $ can be written
as%
\begin{align}
\Phi_{0}\left(  t\right)   &  =X_{0}\left(  t\right)  \cdot\phi_{0}+\int
_{0}^{t}ds\ X_{0}\left(  t-s\right)  \ y\left(  s\right)
,\label{init_cond_1st_step}\\
X_{0}\left(  t\right)   &  =\exp\left(  A_{0}t\right) \nonumber
\end{align}
With this initial condition the solution for $\left(  \ref{step_1_DDE}\right)
$ can be written in terms of its fundamental solution as%
\begin{align}
\Phi_{1}\left(  t\right)   &  =X_{1}\left(  t-\tau_{1}\right)  \ \Phi
_{0}\left(  \tau_{1}\right)  -\int_{0}^{\tau_{1}}ds\ X_{1}\left(  t-s-\tau
_{1}\right)  \ A_{1}\ \Phi_{0}\left(  s\right) \label{init_cond_2nd_step}\\
&  +\int_{\tau_{1}}^{t}ds\ X_{1}\left(  t-s\right)  \ y\left(  s\right)
.\nonumber
\end{align}
$\Phi_{1}\left(  t\right)  $ can be considered as the initial condition for
the 2nd step of iteration, the result of which is the initial condition
defined on the segment $\left[  \tau_{1,}\tau_{2}\right]  .$ The general form
of iteration is easily deduced. Namely,%
\begin{align}
\Phi_{k}\left(  t\right)   &  =X_{k}\left(  t-\tau_{k}\right)  \ \Phi
_{k-1}\left(  \tau_{k}\right)  -\int_{\tau_{k-1}}^{\tau_{k}}ds\ X_{k}\left(
t-s-\tau_{k}\right)  \ A_{k}\ \Phi_{k-1}\left(  s\right)
\label{linear_iteration_loading}\\
&  +\int_{\tau_{k}}^{t}ds\ X_{k}\left(  t-s\right)  \ y\left(  s\right)
.\nonumber
\end{align}
with the final step resulting in the initial condition%
\begin{equation}
\Phi_{L}\left(  t\right)  =\left\{  \Phi_{k}\left(  t\right)  ,\ t\in\left[
\tau_{k},\tau_{k+1}\right]  ,k=0,...,L-1\right\}  . \label{initial_condition}%
\end{equation}
Next note that\ to enable its periodic extension the initial condition has to
be periodic, since it has to be defined on a closed segment. Hence we obtain a
constraint of the form%
\begin{equation}
\Phi_{L}\left(  \tau_{L}\right)  =\Phi_{L}\left(  0\right)  \equiv\phi_{0}.
\label{phi_periodicity}%
\end{equation}
This condition can be used to eliminate an unknown parameter $\phi_{0}$ and
make trajectory loading unique. For example, when there is only one delay we
obtain%
\begin{equation}
\phi_{0}=\left(  \exp\left(  -A_{0}\tau_{L}\right)  -1\right)  ^{-1}\int
_{0}^{\tau_{L}}ds\exp\left(  -A_{0}s\right)  \ y(s).
\label{inhomo_period_constraint}%
\end{equation}
When multiple delays are present $\phi_{0}$ can be determined similarly. The
null space of the mapping $I_{L}$ is also of interest. It describes all the
trajectories that cannot be loaded into the periodic extensions of the initial
conditions. For the example with only one delay this space consists of all
functions with the Laplace transform satisfying%
\begin{equation}
\hat{y}\left(  \lambda\right)  +\phi_{0}=0 \label{1_delay_init_cond}%
\end{equation}
For arbitrary set of delays the null space still forms a one-parameter family
and, therefore, its existence excludes only a very small set of all
trajectories from loading.

Obviously, the detailed loading procedure described above works only for the
linear DDE. However the same principle can be applied to the nonlinear DDE as
well. If the delays are known, then one can proceed with the same iterative
scheme by numerical integration. We speculate that some sort of the analog
integration version of the procedure above might be used in the brain to
implement the loading. With this in mind we can interpret the time $\tau_{L}$
as the ''attention span'' of the population of the neurons that encode a
particular time-dependent pattern. For neurons with a large number of synapses
and, hence a large number of random delays, a continuous version \ of the
iteration in $\left(  \ref{linear_iteration_loading}\right)  $ can be easily written.

\section{Periodic Extensions of the Initial Conditions}

Having shown how trajectories can be loaded into the initial conditions of the
linear DDE, we now construct solutions of the DDE that are the periodic
extension of their initial conditions. First we consider the linear case and
then discuss the modifications added by the presence of nonlinearities.

\subsection{Linear DDEs}

Consider a Fourier series representation of a space of periodic and continuous
initial conditions on $\left[  0,\tau_{L}\right]  $%
\begin{align}
\phi\left(  t\right)   &  =\sum_{n=-Q}^{Q}\exp\left(  i\rho_{n}t\right)  \cdot
b_{n},\;t\in\left[  0,\tau_{L}\right]  ,\label{Fourier_init_cond}\\
\rho_{n}  &  =\frac{2\pi}{\tau_{L}}n,\nonumber
\end{align}
where $b_{n}$ are fixed $N$-dimensional complex-valued vectors such that under
complex conjugation$\ \ \bar{b}_{n}=$ $b_{-n}$ (to ensure that $\phi\left(
t\right)  $ is real-valued). If the initial condition does not belong to this
space we shall consider $\left(  \ref{Fourier_init_cond}\right)  $ as an
approximation of the initial condition with the error of approximation
determined by the truncation of the infinite Fourier series that represents it
at the ''$Q"$th term of the expansion.

Let us extend the domain of definition of the initial condition $\phi\left(
t\right)  $ to all $t\geq0$ by treating $\phi\left(  t\right)  $ as a
continuous periodic function for $t\geq0$ with $\phi\left(  t+\tau_{L}\right)
=\phi\left(  t\right)  .$ Assume\ now that all delays $\tau_{k}$ are fixed but
we are free to vary the matrices $A_{k}.$ The problem of \ finding periodic
extensions is then to find such $A_{k}$ that for $t\geq0$ the function
$\ \phi\left(  t\right)  $ is a solution of the homogeneous equation%
\begin{equation}
\dot{\phi}\left(  t\right)  =\sum_{k=0}^{L}A_{k}\cdot\phi\left(  t-\tau
_{k}\right)  . \label{linear_Homo_DDE}%
\end{equation}
After the substitution of $\left(  \ref{Fourier_init_cond}\right)  $ into
$\left(  \ref{linear_Homo_DDE}\right)  $ we obtain a set of linear equations
on matrices $A_{k}$%
\begin{equation}
\left(  i\rho_{n}I-\sum_{k=0}^{L}A_{k}\cdot\exp\left(  -i\rho_{n}\tau
_{k}\right)  \right)  \cdot b_{n}=0,\ n=0,...,Q. \label{eigenvalue_DDE}%
\end{equation}
Note that the equations for $n=-Q,...,-1$ are redundant, since $A_{k}$ are
real and hence the additional equations can be obtained from $\left(
\ref{eigenvalue_DDE}\right)  $ by complex conjugation. Altogether, because the
$n=0$ equation is real, $\left(  \ref{eigenvalue_DDE}\right)  $ is equivalent
to $\left(  2Q+1\right)  N$ real equations on\ $\left(  L+1\right)  N^{2}$
elements of $A_{k},$ provided that all elements of $A_{k}$ are generically
non-zero. In the case a delay $\tau_{k}$ is given by a random value of delays
in propagation from $"i"$th to $"j"$ neurons, with probability one only one of
the elements of $A_{k}$ is non-zero (see the discussion in the Introduction).
However, in such a case, if $L^{\prime}+1$ is the number of $N\times N$ delay
matrices, then $L+1=\left(  L^{\prime}+1\right)  N^{2}$ and, therefore, the
count of unknowns is the same. We conclude that $\left(  \ref{eigenvalue_DDE}%
\right)  $ can have a unique solution for $A_{k\text{ }}$only when $\left(
2Q+1\right)  =\left(  L+1\right)  N.$ Since $Q$ determines the number of terms
in the Fourier expansion and, therefore, the accuracy of representation of an
arbitrary initial condition, this relation means that for a given number $L$
of delays and $N$ the number of neurons involved in modeling one can achieve
only limited accuracy of representation. Namely, the number of expansion terms
$Q$ is bounded so that $\left(  2Q+1\right)  \leq\left(  L+1\right)  N.$

The system $\left(  \ref{eigenvalue_DDE}\right)  $ can be solved by noticing
that it implies that vectors $b_{n}$ are eigenvectors of linear combinations
of $A_{k}$ with eigenvalues $i\rho_{n}.$ Unlike the typical eigenvalue problem
where $A_{k}$ are known and one needs to find $\rho_{n}$ and $b_{n}$, here the
situation is reversed: one needs to find $A_{k}$ assuming that $\rho_{n}$ and
$b_{n}$ are known. If an eigenvalue $\lambda$ and an eigenvector $b$ are given
then all the matrices $D$ for which they solve the eigenvalue problem can be
written as%
\[
D=\lambda I+B\left(  I-\left\|  b\right\|  ^{-2}b\otimes\bar{b}\right)  ,
\]
where $B$ is arbitrary $N\times N$ matrix and $\left\|  b\right\|  ^{2}%
=\sum_{i=1}^{N}\left|  b_{i}\right|  ^{2}$ and $b\otimes\bar{b}$ is the matrix
that is the outer product of $b$ with its complex conjugate. For given
eigenvalue and eigenvector matrix $D$ has $N^{2}-N$ free parameters.
Therefore, an equivalent way to write $\left(  \ref{eigenvalue_DDE}\right)  $
is%
\begin{align}
\sum_{k=0}^{L}R_{nk}\cdot A_{k}  &  =i\rho_{n}I+B_{n}\left(  I-\left\|
b_{n}\right\|  ^{-2}b_{n}\otimes\bar{b}_{n}\right)
,\ n=0,...,Q,\label{A_explicit}\\
R_{nk}  &  =\exp\left(  -i\rho_{n}\tau_{k}\right)  , \label{def_R}%
\end{align}
where $B_{n}$ is an $N\times N$ complex matrix, which effectively has
$N^{2}-N$ free parameters. This linear system of equations always has
solutions for appropriate choice of parameters $Q,L,N.$ For some choices the
solution is unique. This concludes the proof of existence of the periodic extensions.

Let us consider $\left(  \ref{eigenvalue_DDE},\ref{A_explicit}\right)  $ for
various choices of parameters. When $N=1$ the second term in the RHS of
$\left(  \ref{A_explicit}\right)  $ vanishes and the dependence on $b_{n}$
drops out. Consequently, the system $\left(  \ref{eigenvalue_DDE}\right)  $
becomes a system of $\left(  2Q+1\right)  $ real linear equations on $L+1$
real parameters $A_{k}$%
\begin{align}
\sum_{k=0}^{L}S_{nk}A_{k}  &  =-\rho_{n},\ n=1,...,Q,\label{sin_N_1}\\
\sum_{k=0}^{L}C_{nk}A_{k}  &  =0,\ \ \ \ \ n=0,1,...,Q,\label{cos_N_1}\\
S_{nk}  &  =\sin\left(  \rho_{n}\tau_{k}\right)  ,\label{def_sin_N_1}\\
C_{nk}  &  =\cos\left(  \rho_{n}\tau_{k}\right)  . \label{def_cos_N_1}%
\end{align}
When $L=2Q,$ provided $\det T\neq0,$ the equations can be solved uniquely by
inversion of the matrix $T$%
\begin{equation}
T=\left(
\begin{array}
[c]{ccc}%
\sin\left(  \rho_{1}\tau_{0}\right)  & ... & \sin\left(  \rho_{1}\tau
_{L}\right) \\
... & ... & ...\\
\sin\left(  \rho_{Q}\tau_{0}\right)  & ... & \sin\left(  \rho_{Q}\tau
_{L}\right) \\
1 & ... & 1\\
... & ... & ...\\
\cos\left(  \rho_{Q}\tau_{0}\right)  & ... & \cos\left(  \rho_{Q}\tau
_{L}\right)
\end{array}
\right)  \label{real_N_1}%
\end{equation}
For arbitrary \ $N$ , $L=2Q,$ and $\det T\neq0$ we can write down the
solutions for $\left(  \ref{eigenvalue_DDE}\right)  $ $R$ as%
\begin{align}
A_{k}  &  =\sum_{k=0}^{2Q}T_{kn}^{-1}\Gamma_{n}%
,\ \ \ \ \ \ \ \ \ \ \ \ \ \ \ \ \ \ \ \ \ \ \ \ \ \ \ \ \ \ \ \ \ \ \ \ \ \ \ \ \ \ \ \ \ \ \ \ \ \ k=0,1,...,L.
\label{complex_QeqL}\\
\Gamma_{n}  &  =\left(  \rho_{n}I+\operatorname{Im}\left(  B_{n}\left(
I-\left\|  b_{n}\right\|  ^{-2}b_{n}\otimes\bar{b}_{n}\right)  \right)
\right)  ,\ \ n=1,...,Q\label{Gamma_sin}\\
\Gamma_{n}  &  =\operatorname{Re}\left(  B_{n}\left(  I-\left\|
b_{n}\right\|  ^{-2}b_{n}\otimes\bar{b}_{n}\right)  \right)
,\ \ \ \ \ \ \ \ \ \ \ \ \ \ \ \ n=0,1,...,Q \label{Gamma_cos}%
\end{align}
Since the matrices $\Gamma_{n}$ contain $\left(  N^{2}-N\right)  $ free
parameters, to obtain a unique solution one needs to increase the value of $Q$
by factor of $\left(  N^{2}-N\right)  $ to constrain the additional degrees of freedom.

Some remarks about the meaning of the periodic extensions should be made.
Since the initial conditions were taken as a linear combination of exponential
solutions, the existence of the extensions is equivalent to the existence of
$2Q+1$ pair-wise conjugate zeroes of the characteristic function that are
located on the imaginary axis. As mentioned in the previous section, there can
be only a finite number of zeroes of the characteristic function in any finite
width vertical strip of the complex plane. Hence, there could be only a finite
number of zeros of the characteristic function lying on the imaginary axis. As
a result, $Q\ must$ be finite and only finite-dimensional spaces of initial
conditions can be extended periodically. A practical consequence of this is
that one can store a trajectory as a periodic extension only approximately,
with the error of approximation given by the error induced by truncating the
Fourier series. However, as follows from the theory of Fourier expansions, the
error can be made arbitrarily small at least in $L^{2}$ norm by increasing $N.$

\subsubsection{Non-linear DDE}

Let us now consider the existence of the periodic extensions in the presence
of nonlinearities. Take, for example, the extended Hopfield model with
dynamics%
\begin{equation}
\dot{x}=\sum_{k=0}^{L}A_{k}\cdot\sigma\left(  x\left(  t-\tau_{k}\right)
\right)  +y(t)
\end{equation}
Proceeding as before we expand the periodic the initial condition in the
Fourier series for $t\in\left[  0,\tau_{L}\right]  $ as in $\left(
\ref{Fourier_init_cond}\right)  $ and obtain that for $\phi\left(  t\right)  $
to be a solution of homogeneous DDE it needs to satisfy%
\begin{equation}
\dot{\phi}\left(  t\right)  =\sum_{k=0}^{L}A_{k}\cdot\sigma\left(  \phi\left(
t-\tau_{k}\right)  \right)  \label{nonlinear_init_cond_ext}%
\end{equation}
Substitution of $\left(  \ref{Fourier_init_cond}\right)  $ into $\left(
\ref{nonlinear_init_cond_ext}\right)  $ and expansion of the nonlinear term in
Fourier series yields a system of equations on the unknown matrices $A_{k}$%
\begin{equation}
\left(  i\rho_{n}b_{n}-\sum_{k=0}^{L}A_{k}\cdot\exp\left(  -i\rho_{n}\tau
_{k}\right)  \cdot\Lambda_{n}\left(  \tau_{k};\left\{  b_{m}\right\}  \right)
\right)  =0,\ n=-Q_{1},...,Q_{1}%
\end{equation}
where $Q_{1}$ is not necessarily equal to $Q$ and for each $p=1,....N$ \ the
coefficient $\left(  \Lambda_{n}\left(  \tau_{k};\left\{  b_{m}\right\}
\right)  \right)  _{p}$ is defined by%
\begin{equation}
\Lambda_{n}\left(  \tau_{k};\left\{  b_{m}\right\}  \right)  =\frac{1}%
{\tau_{L}}\int_{0}^{\tau_{L}}dt\exp\left(  -i\rho_{n}\ \left(  t-\tau
_{k}\right)  \right)  \ \sigma\left(  \sum_{m=-Q}^{Q}\exp\left(  i\rho
_{m}\ \left(  t-\tau_{k}\right)  \right)  \cdot b_{m}\right)
\label{lambda_def}%
\end{equation}
The additional time-independent factor $\exp\left(  i\rho_{n}\ \tau
_{k}\right)  $ in the definition of $\Lambda_{n}\left(  \tau_{k};\left\{
b_{m}\right\}  \right)  $ ensures that if $\sigma\left(  z\right)  =z$ then
$\Lambda_{n}\left(  \tau_{k};\left\{  b_{m}\right\}  \right)  =b_{n}.$

Now the\ vector coefficients of the Fourier expansion $\left(
\ref{Fourier_init_cond}\right)  $ cannot be factored out and the solution for
$A_{k}$ will depend not only on the delays $\tau_{k}$ but also on $b_{n}$even
for $N=1$. Such dependence is desirable in a memory model, since it implies
that the choice of the parameters $A_{k}$ is pattern specific.

When $Q_{1}$ $\neq$ $Q$ and $Q_{1}<\infty$ we obtain essentially the same
system of equations as before. One class of nonlinearities when this occurs is
a set that may be called polynomial nonlinearities with%
\begin{equation}
\sigma\left(  z\right)  =\sum_{N_{1}}^{N_{2}}c_{n}z^{n},\ 0<N_{1}\leq
N_{2}<\infty\label{polynomial nonlinearity}%
\end{equation}
For example, if we assume that all trajectories that we wish to store are
bounded, in order to introduce a sigmoid nonlinearity that is frequently used
in memory models, one can take a cubic nonlinearity with
\begin{equation}
\sigma\left(  z\right)  =-\left(  1/3\right)  \alpha^{2}z^{3}+\beta
^{2}z,\ \alpha,\beta>0 \label{cubic nonlinearity}%
\end{equation}
and load only such trajectories for which $\sigma\left(  z\right)  \leq\left(
2/3\right)  \left(  \beta^{3}/\alpha\right)  .$ Substitution of $\left(
\ref{cubic nonlinearity}\right)  $ into $\left(  \ref{nonlinear_init_cond_ext}%
\right)  $ results in
\begin{align}
\left(  i\rho_{n}b_{n}-\sum_{k=0}^{L}A_{k}\cdot\exp\left(  -i\rho_{n}\tau
_{k}\right)  \cdot\left(  \beta^{2}b_{n}+\frac{1}{3}\sum_{p,q=-Q}^{Q}%
b_{p}b_{q}b_{n-p-q}\right)  \right)   &  =0,\ \\
n  &  =-Q_{1},...,Q_{1}\nonumber
\end{align}

As in the linear case this equation can be still considered as an eigenvalue
problem for $b_{n}$, except for a nonlinear operator $\tilde{A}=g\left(
A\right)  .$ On the other hand this equation can also be considered as a
linear equation on $A_{k}$ for a given set of $b_{n}.$ One novel feature that
appears in the nonlinear case is that since we have a cubic nonlinearity, in
general, up to three different values for each $b_{n}$ are possible so that
the term $\left(  \beta^{2}b_{n}+\frac{1}{3}\sum_{p,q=-Q}^{Q}b_{p}%
b_{q}b_{n-p-q}\right)  $ has the same value and hence the solution for $A_{k}$
is the same. This observation can be used for multiple memory storage of a set
of $M$ desired memories $\left\{  b_{n}^{B}\right\}  ,B=1,...,M.$ Consider the
following system of equations%
\[
\left(  \beta^{2}b_{n}+\frac{1}{3}\sum_{p,q=-Q}^{Q}b_{p}b_{q}b_{n-p-q}\right)
=C_{n},\ n=-Q_{1},...,Q_{1}%
\]
This cubic system of $2Q+1$ equations has $3^{2Q+1}$ choices of solutions. As
a result, in general, one can load $3^{2Q+1}$ memories corresponding to the
same choice of matrices $A_{k}.$ The same argument applies to any polynomial
nonlinearity defined by $\left(  \ref{polynomial nonlinearity}\right)  $ with
the corresponding maximum number of memories growing as $N_{2}^{2Q+1}.$ Of
course for this storage method to be practical a prescription should be given
for a way to store a given set of memories for a given nonlinearity, rather
then choosing a nonlinearity to fit the needed set of memories as we described
above. In addition a thorough investigation of the basins of attraction of the
multiple memories needs to be carried out. These and other questions shall be
addressed in a future publication.

A problem can arise if $Q_{1}$ is sufficiently large, that is the system can
become overdetermined if $Q_{1}$ is large enough. If $Q_{1}$ is finite the
problem can be cured by choosing appropriately large $L.$ The real problem
appears when nonlinearity is such that $Q_{1}=\infty.$ Then, one would expect
that the system $\left(  \ref{nonlinear_init_cond_ext}\right)  $ generically
has no solutions. When $Q_{1}=\infty$ our approach can be nevertheless applied
approximately in the following sense. If $\Lambda_{n}\left(  \tau_{k};\left\{
b_{m}\right\}  \right)  \rightarrow0$ sufficiently fast as $n\rightarrow
\infty,$ we can truncate the system at a finite $Q_{\max}$ and the proceed as
before with computing the $A_{k}$. Of course since we zeroed the coefficients
$\Lambda_{n}\left(  \tau_{k};\left\{  b_{m}\right\}  \right)  $ for
$n>Q_{\max}$ in effect we substituted the original nonlinearity $\sigma\left(
z\right)  $ with another one $\tilde{\sigma}\left(  z\right)  $ and the
periodic extensions obtained for $\tilde{\sigma}\left(  z\right)  $ will not
be periodic solutions for the original nonlinearity. However if the original
nonlinear DDE is stable with regard to small perturbations of the solutions,
the aperiodicity will not grow in time and will remain small. Therefore we
still be able to store time-variable patterns, although with an additional error.

The criteria for the stability of the periodic extension follow from the
stability analysis of the associated DDE. For the nonlinear case the stability
of the nonlinear DDE is equivalent to the stability of the linearized version
of the original DDE. The basic result about the stability of a linear DDE is
that any solution of a DDE is stable if and only if the largest real part of
the roots of the characteristic equation is non-positive. The criteria for the
roots of the characteristic equation to lie in the complex left half-plane
have been studied extensively. A summary of the results can be found in
\cite{Hale}.

\section{Parameter Adaptation Dynamics}

If the construction of periodic extensions is realized in the physiology of
the brain then there must exist an algorithm that adjusts weights and/or
delays to arrive at the needed values so that a population of neurons learns
the appropriate weights after many repetitions of exposure to the external
input. In this section we present examples of such learning algorithms. Of
course we do not claim that either of the algorithm is actually implemented in
the brain. Neither is learning a central point of this paper. Nevertheless it
is instructive to consider some possibilities to estimate the difficulty of
the problem.

We shall consider that the delays are known and fixed and only the weight
matrices $A_{k}$ need to be determined. The delays can also be considered as
variables instead or together with $A_{k}$ . Investigation of this possibility
we leave for elsewhere.

A learning algorithm is not difficult to construct by introducing a Liapunov
dynamics in the space of weights so that at the fixed point of the evolution
we obtain the needed periodic extension. The weight dynamics can be
consequently constructed by defining an error functional that measures the
distance from the desired solution to the initial condition and using it as a
Liapunov function. Consider, for example, ''slow'' learning done during the
multiple presentations - ''epochs'' - of the same initial condition. The
$"m"$th update of $A_{k}$ would then be done once per ''epoch'': one update
for each time interval $\left[  m\ \tau_{L},\left(  m+1\right)  \ \tau
_{L}\right]  .$ As an example we can take the $"m"$th epoch error functional
as%
\begin{equation}
V_{m}\left[  A\right]  =\int_{m\ \tau_{L}}^{\left(  m+1\right)  \tau_{L}%
}dt\ \left|  x\left(  t\right)  -\phi_{0}\left(  t\right)  \right|  ^{2},
\label{DDE_Liapunov}%
\end{equation}
where $x\left(  t\right)  $ is the solution of the DDE with a given initial
condition $\phi_{0}\left(  t\right)  .$ ''Fast'' adaptation within the single
epoch can also be defined in a similar way, although from the point of view of
biology it might be more suitable for adaptation of delays. The error
functional $\left(  \ref{DDE_Liapunov}\right)  $\ is phenomenologically
appealing since both $x\left(  t\right)  $ and $\phi\left(  t\right)  $ can be
physically ''measured'' by the encoding population of neurons. The discrete
evolution in the space of weights with fixed delays can then be written as
\begin{equation}
\left(  A_{k}\right)  _{m+1}=\left(  A_{k}\right)  _{m+1}-\varepsilon
\frac{\partial}{\partial A_{k}}V_{m}\left[  A\right]  , \label{A-dynamics}%
\end{equation}
where $\varepsilon$ is a small parameter. If the fundamental solution can be
computed then $x\left(  t\right)  $ in $\left(  \ref{DDE_Liapunov}\right)  $
is given by%
\begin{equation}
x\left(  t\right)  =X\left(  t-\tau_{L}\right)  \phi\left(  \tau_{L}\right)
+\sum_{k=1}^{L}\int_{\tau_{L}-\tau_{k}}^{\tau_{L}}d\tau\;X\left(  t-\tau
-\tau_{k}\right)  \;A_{k}\;\phi\left(  \tau\right)  ,\nonumber
\end{equation}
where the dependence of the fundamental solution $X(t)$ on the weights is
given by (\ref{fund_sol},\ref{char_eq}). Computation of the appropriate
derivatives of $V[A]$ involves computing the $A-$derivatives of the
fundamental solution. These can be written out as%
\begin{equation}
\frac{\partial X_{pq}}{\partial\left(  A_{k}\right)  _{ij}}=\frac{1}{2\pi
i}\int_{\left(  c\right)  }d\lambda\;\exp\lambda\left(  t-\tau_{k}\right)
\left(  H^{-2}\left(  \lambda\right)  \right)  _{pi}\ \delta_{qj},
\label{X_A_deriv}%
\end{equation}
which enables us to compute\ $\frac{\partial}{\partial A_{k}}V_{m}\left[
A\right]  $.

A more direct approach to learning the weights is to use the characteristic
equation itself. We know from the discussion in the preceding sections that
matrices $A_{k}$ must be chosen in such a way that the characteristic equation
has purely imaginary roots located at $\lambda_{n}=\frac{2\pi i}{\tau_{L}}n,$
$n=-Q,...,Q.$ Therefore an alternative Liapunov function for the weight
dynamics can be chosen as
\begin{align}
V_{H}\left[  A\right]   &  =\frac{1}{2}\sum_{n=-Q}^{Q}\left|  \det H\left(
\lambda_{n}\right)  \right|  ^{2}\label{char_Liapunov}\\
H\left(  \lambda\right)   &  =\lambda I-\sum_{k=0}^{L}A_{k}\exp\left(
-\lambda\tau_{k}\right) \nonumber
\end{align}
The weight dynamics induced by this Liapunov function is given by%
\begin{equation}
\dot{A}_{k}=-\varepsilon\operatorname{Re}\left(  \sum_{n=-Q}^{Q}\exp\left(
-\lambda_{n}\tau_{k}\right)  \det\bar{H}\left(  \lambda_{n}\right)
H^{-T}\left(  \lambda_{n}\right)  \right)  \label{char_Liap_A_dyn}%
\end{equation}
where bar denotes complex conjugation. This system of equations conceptually
simpler than the one obtained from $\left(  \ref{DDE_Liapunov}\right)  ,$
since it does not involve integration over time.

Yet another approach is to use the explicit form of the solutions for $A_{k}$
and define the norm in the space of matrices by%
\begin{align}
V_{Y}\left[  A\right]   &  =\frac{1}{2}\sum_{k=0}^{L}tr\left(  A_{k}%
-Y_{k}\right)  \left(  A_{k}-Y_{k}\right)  ^{T}\\
Y_{k}  &  =\sum_{k=0}^{2Q}T_{kn}^{-1}\Gamma_{n}\nonumber
\end{align}
where $T,\Gamma$ are defined by $\left(  \ref{def_sin_N_1}\ref{def_cos_N_1}%
,\ref{Gamma_sin},\ref{Gamma_cos}\right)  .$ Differentiation with respect to
unknown $A_{k}$ results in%
\[
\dot{A}_{k}=-\varepsilon\left(  \left(  A_{k}-Y_{k}\right)  \right)
\]

The three algorithms we presented all apply only to the linear case. For
applications in biology the DDE are typically nonlinear. In this case one can
use a stochastic annealing approach to the error functional $V_{m}\left[
A\right]  $ defined by $\left(  \ref{DDE_Liapunov}\right)  .$ Another
alternative is to restrict learning to the linear regime of the nonlinear DDE.
This is possible because the typical nonlinearities in memory models, e.g. the
Hopfield model, do have a region of definition of the nonlinearity where
$\sigma\left(  z\right)  \approx z.$

\section{Summary}

In this paper we presented a method for approximate storage of the
trajectories of nonlinear dynamical systems as periodic or almost periodic
solutions of the linear and nonlinear delay-differential equations.

Although the main motivation for this work was to develop a model for
time-variable pattern storage by the brain, the results might have wider
applicability. Indeed, they indicate that linear delay differential equations
are, in a sense, universal models for nonlinear dynamical systems. Choosing
appropriately large $N$ one in principle can store arbitrary (but finite)
number of periodic orbits of a nonlinear dynamical system.

In addition to providing a method for storage of time-dependent patterns our
model has other features that are attractive for a biological interpretation.
For example, the method requires global synchronization of populations of
neurons, since the initial conditions have to be represented via Fourier
series with the same fundamental frequency $\frac{2\pi}{\tau_{L}}$. The
maximum delay $\tau_{L}$ can be thought of the attention span of the
population: periodic trajectories with period larger then $\tau_{L}$ cannot be
stored. The model does not have unknown parameters. Once the delays are known
the weights can be determined uniquely for a given trajectory of the environment.

There are a number of open questions about our method. Although the situation
with the linear DDE seems to be clear, a more detailed investigation of the
effect of the presence of nonlinearity should be carried out. One beneficial
effect of nonlinearity could be multistability for a chosen set of the
coefficients $A_{k},\tau_{k}$: the existence of different, initial condition
dependent periodic solutions similar to the existence of multiple fixed points
in the Hopfield model. This would make multiple trajectory storage more
efficient: a linear DDE can store additional orbits only by increasing its
dimension linearly with the number of the orbits.

We have shown that nonlinearity brings along one interesting feature:
multistability. More than one periodic solution can be stored for a given set
of matrices $A_{k}.$ When nonlinearity is polynomial, one can estimate that
maximum possible number of memories grows exponentially in the number of
Fourier coefficients. Additional indication that the multistability should
exist comes from considering a continuum limit of $\left(
\ref{nonlinear_init_cond_ext}\right)  ,$ a limit that is reasonable to take
when there are large number of delays that depend smoothly on their index.
This is the situation when neurons in the brain have large dendritic trees.
With obvious definitions the continuum limit can be formally written as%
\begin{equation}
\dot{\phi}\left(  t\right)  =\lambda\int_{0}^{1}ds\ A\left(  s\right)
\cdot\sigma\left(  \phi\left(  t-\tau\left(  s\right)  \right)  \right)
\label{Hammerstein_integral_eq}%
\end{equation}
for $\lambda$ a parameter and for some monotonic delay function $\tau\left(
s\right)  .$ Here we normalized the matrix-valued function $A\left(  s\right)
$ so that $tr\left(  A\left(  s\right)  ^{T}A\left(  s\right)  \right)  =1.$
With the use of the Fourier transform this equation can be related to the
Volterra and Hammerstein classes of nonlinear integral \ equations, which
exhibit bifurcations in the parameter $\lambda$. The theory of nonlinear
integral equations has been exhaustively studied and a number of \ criteria
are available for determining the bifurcation points \cite{Yosida}. These
depend mainly on the growth properties of the nonlinearity $\sigma\left(
z\right)  $. A more detailed analysis of\ $\left(
\ref{Hammerstein_integral_eq}\right)  $ shall be considered elsewhere.

\begin{acknowledgement}
I would like to thank Holger Kantz and members of the Nonlinear Time Series
Analysis Group at MPI-PKS for useful discussions.
\end{acknowledgement}

\end{document}